\documentclass[bibyear]{aa} 

\usepackage{natbib}
\usepackage{color}
\usepackage{graphicx}
\usepackage{ae,aecompl}
\usepackage{txfonts}

\newcommand{\JF}[1]{{\textbf{\textcolor{magenta}{[JF: #1]}}}}

\begin{document}

   \title{Simulations of cluster ultra-diffuse galaxies in MOND}
\authorrunning{S.T. Nagesh, J. Freundlich, B. Famaey et. al.}

   \author{Srikanth T. Nagesh
          \inst{1}  
          \and
          Jonathan Freundlich\inst{1}
          \and 
          Benoit Famaey\inst{1}
          \and 
          Michal B\'ilek\inst{2,5} 
                    \and 
          Graeme Candlish\inst{3}
                    \and \\
          Rodrigo Ibata\inst{1}
                     \and 
          Oliver M\"uller\inst{4}
          }

   \institute{
        Universit\'e de Strasbourg, CNRS, Observatoire astronomique de Strasbourg, UMR 7550, F-67000 Strasbourg, France\\
        \email{togerenagesh@unistra.fr}
        \and
        LERMA, Observatoire de Paris, CNRS, PSL Univ., Sorbonne Univ., 75014 Paris, France
        \and
        Instituto de F\'isica y Astronom\'ia, Universidad de Valpara\'iso, Gran Breta\~ na 1111, Valpara\'iso, Chile
        \and 
        Institute of Physics, Laboratory of Astrophysics, Ecole Polytechnique Fédérale de Lausanne (EPFL), 1290 Sauverny, Switzerland
        \and
        Scottish Universities Physics Alliance, University of Saint Andrews, North Haugh, Saint Andrews, Fife, KY16 9SS, UK
             }

   \date{Received XXX; accepted YYY}

 
  \abstract
  {Ultra-diffuse galaxies (UDGs) in the Coma cluster have velocity dispersion profiles that are in full agreement with the predictions of Modified Newtonian Dynamics (MOND) in isolation. However, the external field effect (EFE) from the cluster seriously deteriorates this agreement. It has been suggested that this could be related to the fact that UDGs are out-of-equilibrium objects whose stars have been heated by the cluster tides or that they recently fell onto the cluster on radial orbits, such that their velocity dispersion may not reflect the EFE at their instantaneous distance from the cluster center. Here, we simulate UDGs within the Coma cluster in MOND, using the Phantom of Ramses 
  (\textsc{por}) code, and show that if UDGs are initially at equilibrium within the cluster, tides are not sufficient to increase their velocity dispersions to values as high as the observed ones. On the other hand, if they are on a first radial infall onto the cluster, they can keep high velocity dispersions without being destroyed until their first pericentric passage. We conclude that, without alterations such as a screening of the EFE in galaxy clusters or much higher baryonic masses than currently estimated, in the MOND context UDGs must be out-of-equilibrium objects on their first infall onto the cluster.}
   
   
   


   \maketitle

%
\section{Introduction}
\label{section:introduction}
The need for an additional component in the matter sector, beyond the one described by the standard model of particle physics, is backed, in the context of General Relativity (GR) and its weak-field Newtonian counterpart, from a plethora of observations at scales ranging from galaxies to the whole observable Universe. However, it had also been suggested four decades ago \citep{Milgrom_1983b,Milgrom_1983,Bekenstein_Milgrom_1984} that, at least on galactic scales, phenomena attributed to this additional matter component could also be attributed in principle to new gravitational degrees of freedom instead of new particles. This idea, known as Modified Newtonian dynamics \citep[MOND, see][for extensive reviews]{Famaey_McGaugh_2012,Milgrom_2014, Banik_Zhao_2022}, postulates that weak-field deviations from Newtonian dynamics occur in systems with accelerations below Milgrom's constant $a_0 \approx 1.2 \times 10^{-10}$ m/s$^{2}$ $\approx$ 3.9 pc/Myr$^{2}$ \citep{Begeman_1991, Gentile_2011, Desmond_2024}. Well below this threshold, and until the external gravitational field dominates over the internal one, the gravitational acceleration would become $g=\sqrt{g_{_N} a_{_0}}$, where $g_{_N}$ is the Newtonian gravitational acceleration. This simple prescription automatically predicts the asymptotic flatness of galaxy rotation curves but also makes several important non-trivial predictions. In particular, it predicts a relation between the total baryonic mass and the asymptotic circular velocity of rotationally-supported disk galaxies, with no dependence of the residuals on the surface density of the disks, a power-law slope of 4, and no change of slope at high masses. This relation, known as the Baryonic Tully-Fisher Relation (BTFR) has been repeatedly confirmed for rotationally-supported galaxies \citep{McGaugh_2000,Lelli2019,Diteodoro_2023}. Even more non-trivially, MOND predicts that BTFR `twins', i.e. disk galaxies sharing the same baryonic mass and asymptotic circular velocity, should display very different rotation curve shapes as a function of surface density. In fact, this is precisely what is observed \citep[e.g.][]{Blok_1997,Swaters_2009}. Interpreted in the dark matter context, this would mean that disk galaxies should display a variety of CDM halo density profiles as a function of the surface density of the baryons, which remains very surprising today in the standard $\Lambda$CDM context~\citep{Oman_2015,Ghari_2019}. In summary, this observed dependence of rotation curve shapes on baryonic surface density, together with the surprising independence of the BTFR on that same baryonic surface density, is the main argument to take MOND seriously as a possible alternative to CDM. This phenomenology is encapsulated into the observational Radial Acceleration Relation for disk galaxies \citep[RAR,][]{McGaugh_2016,Lelli_2017,Stiskalek}, which connects the radial dynamical acceleration inferred from kinematics with that predicted from the observed baryonic distribution. 

Stellar systems that have low internal gravitational accelerations ($g \ll a_0$) are in principle an ideal testing ground for MOND. Ultra-diffuse galaxies \citep[UDGs;][]{Fosbury1978, Sandage1984, Karachentsev2000} are low surface brightness (LSB) objects with a typical central surface brightness $\mu_{g,0}$ > 24 mag/arcsec$^2$, optical luminosities ranging from $10^7 - 10^8 \rm ~L_\odot$, and large effective radii (as compared to other dwarfs) $R_{\rm eff} > 1.5$ kpc, such that their internal accelerations are very low. UDGs have been observed both in the field \citep{Leisman2017,Roman2017, Prole2019, Bautista_2022}, and in galaxy groups and galaxy clusters \citep{van_Dokkum_2015a, vanDokkum2015b, Janowiecki_2015, Mihos2015, Mihos2017, Yagi_2016, Koda_2015, Martinez_2015, Venhola2017, Muller_2018, Marleau_2021}. For example, in the Coma cluster, there are $\sim 10^3$ detected UDGs \citep[e.g.][]{Bautista_2023}. Multiple scenarios for their formation in the standard $\Lambda$CDM context have been proposed \citep[e.g.][]{vanDokkum2015b, Van_Dokkum_2016, Amorisco2016, Beasley2016b, DiCintio2017, Greco2018, Toloba2018, Jiang2019, Freundlich2020a, Freundlich2020b}, but no consensus has been reached, and a large uncertainty over their dark matter content still persists \citep{Van_Dokkum_2016, Van_Dokkum_2018, Van_Dokkum_2019a, Wasserman2019, Nusser2019, Emsellem2019, Haslbauer_2019, Muller2021}.

In the MOND context, \citet{Freundlich_2022} investigated a sample of 11 UDGs \citep{van_Dokkum_2015a,vanDokkum2016,vanDokkum2017,vanDokkum2019b,Chilingarian2019} with measured stellar velocity dispersion profiles in the Coma cluster and noted that those UDGs seem to be in-line with the MOND prediction if these galaxies were isolated \citep[see also][]{Bilek_2019, Haghi_2019_DF44}. However, the non-linear nature of MOND gravity should imply that the dynamics of a system is regulated by the total gravitational field (both its internal field $g$ and the {\it external} one $g_e$ in which it is embedded). If $g < g_e$, as is the case for UDGs in the Coma cluster, the system should experience an `external field effect' \citep[EFE;][]{Milgrom_1983b,Bekenstein_Milgrom_1984,Famaey_McGaugh_2012, McGaugh_Milgrom_2013} which would damp the 
rotational velocities or velocity dispersions compared to those predicted by MOND in isolation \citep[e.g.,][]{McGaugh_Milgrom_2013,Pawlowski_2015, Hees_2016, Famaey_2018, Kroupa_2018_Nature, Bilek_2018, Haghi_2019_DF44, Muller_2019,Chae_2020,Chae_2021, Oria_2021}. The EFE is also an observational necessity in the MOND context to explain certain phenomena like the escape velocity curve of the Milky Way \citep{Famaey_2007, Banik_2018_escape,Oria_2021}. 
Therefore, UDGs inside clusters should be entirely EFE-dominated in the MOND context, meaning that the result of \citet{Freundlich_2022} seems to either (i) contradict MOND or (ii) could mean that the EFE is screened inside the Coma cluster for some deep theoretical reasons related to the yet-to-be-found fundamental theory underpinning the MOND paradigm. However, in the context of classical modified gravity theories of MOND, other possible explanations might be that UDGs are out-of-equilibrium objects (iii) whose stars have been heated by the cluster tides or (iv) that recently fell onto the cluster on radial orbits, such that their velocity dispersion may not reflect the EFE at their instantaneous distance from the cluster center.

The present work focuses on testing these two last hypotheses (iii) and (iv), via detailed $N-$body simulations using the \textsc{por} patch of the \textsc{ramses} code. The article is structured as follows: Section 2 describes the numerical methods as well as the simulations setups, Section 3 discusses the results of the simulations and Section 4 concludes.

\section{Methods}
\label{sec:methods}

MOND can in principle be formulated as a modification of Newton's second law, but such formulations cannot be considered as fully-fledged theories yet \citep{MI_1994, MI_2022}. On the other hand, theories based on adding new gravitational degrees of freedom to GR have been well-developed over the last four decades, including in scalar-tensor form \citep{Bekenstein_Milgrom_1984} and later in tensor-vector-scalar form to account for gravitational lensing \citep{Bekenstein2004}, their latest versions even managing to reproduce cosmological observables in the linear regime of structure formation \citep{Skordis2020,Blanchet2024}. Such theories are typically calibrated to reproduce a generalised classical Lagrangian for gravity in the weak-field limit, associated to a non-linear MOND Poisson equation. Two main such classical Lagrangians have been proposed, one called the aquadratic Lagrangian (AQUAL) theory, developed by \citet{Bekenstein_Milgrom_1984}, and the other called quasi-linear MOND (QUMOND) developed by \citet{QUMOND}. These formulations enable one to apply MOND to systems that deviate from spherical symmetry, where the algebraic relation $g=\sqrt{g_{_N} a_{_0}}$ in the weak-field regime cannot be exact \citep[see, e.g.,][]{Bekenstein_Milgrom_1984, Brada_1995,Famaey_McGaugh_2012}.

Both these classical formalisms have been numerically implemented and tested on diverse scenarios. 
For example, AQUAL was implemented in a \textit{N}-body code developed by \citet{Brada_1999} which was used to study the stability of disk galaxies, and \citet{Tiret_2008} later developed a multi-grid Poisson solver to study the evolution of spiral galaxies using pure stellar disks and gas dynamics using a sticky particle scheme \citep{Tiret_2008_gas}. The $N$-mody code \citep{Londrillo_2009} was also developed in order to study dynamical questions such as the radial orbit instability in the AQUAL context \citep{Nipoti_2011}.
Two main $N$-body and hydrodynamical codes have been developed as patches of the adaptive mesh refinement (AMR) code \textsc{ramses} \citep{Teyssier_2002}. \textsc{ramses} is equipped with a Newtonian Poisson solver for gravitational computations, and a second-order Godunov scheme with a Riemann solver for the Euler equations, which allows one to run both  $N-$body and hydrodynamical simulations with star formation. The \textsc{raymond} patch has both AQUAL and QUMOND implemented, and has for instance been used to run cosmological simulations \citep{Candlish_2015}. The phantom of ramses \citep[\textsc{por} patch,][]{Lughausen_2015, Nagesh_2021} is a publicly available patch\footnote{The \textsc{por} package, extraction software, and other relevant algorithms are available at \url{bitbucket.org/SrikanthTN/bonnPoR/src/master/}, along with a \textsc{por} manual to setup, run, and analyse isolated disk galaxy simulations in MOND \citep{Nagesh_2021}\label{bitbucket}.}  numerically implementing the QUMOND Poisson equation within the \textsc{ramses} Poisson solver, that has been widely used over the last decade to test QUMOND predictions in a plethora of systems \citep{Lughausen_2013,Thomas_2017,Thomas_2018,Bilek_2018,Bilek_2022b,Renaud_2016,Banik_2020_M33,Wittenburg_2020,Eappen_2022,Banik_2022_satellite_plane,Nagesh_2023, Wittenburg_2023}. 
Several other independent codes have also been used to test cosmology in the context of MOND \citep{Llinares_2008, Angus_2011, Angus_2013}.

The simulations presented here are carried out using \textsc{por}. With MOND gravity turned on, the field equation for the gravitational potential $\Phi$ reads as
\begin{equation}
    \nabla^2 \Phi ~\equiv~ - \nabla \cdot \vec{g}  ~=~ - \nabla \cdot \left( \nu \vec{g}_{_N} \right) \, ,
    \label{eq:poisson}
\end{equation}
where $\vec{g}_{_N}$ and $\vec{g}$ are the Newtonian and MONDian gravitational acceleration vectors respectively.
The function $\nu$ has $g_{_N}/a_0$ as argument, and is the MOND interpolating function that dictates the transition between Newtonian and MONDian regimes, for which we use the so-called `simple' form \citep{Famaey_Binney_2005,Famaey2012}. At each step, \textsc{por} computes $\vec{g}_{_N}$ from the baryon density $\rho_{_b}$ by solving the standard Poisson equation, then uses the interpolating function to compute the new source term on the right-hand side of Eq.~(\ref{eq:poisson}), and solves the standard Poisson equation a second time to find the QUMOND potential $\Phi$.

The UDGs used in our simulations are initially modelled as S\'ersic spheres, following \citet{Bilek_2022b}.
We assume a total mass $M_{\textrm{UDG}} = 6 \times 10^7 ~ \textrm{M}_\odot$, an effective radius $R_{\textrm{eff}} = 1.5$ kpc, and a S\'ersic index $n = 1$. These parameters are approximately chosen to be the median of the observed UDG sample analyzed in \citet{Freundlich_2022}, with exception of DF44 and DFX1. 
We de-project the two-dimensional S\'ersic light profile using the semi-analytical approximation proposed by \citet{Lima_Neto1999}, with an update from \citet{Marquez2000}, cf. \citet[][Section 3.1.2]{Freundlich_2022}, numerically invert the corresponding enclosed mass profile, and sample positions of particles from the resulting inverted cumulative distribution function. 
%
%
%
For a particle at a given radius, the speed $v$ is drawn from a Gaussian distribution with standard deviation given by the velocity dispersion, which is obtained from the Jeans equation \citep[Eq.~4.125 of ][]{Binney_Tremaine2008} assuming isotropy. 
%
%
Random numbers are drawn from a ${\cal N}(0,1)$ Gaussian distribution for $v_x, v_y, v_z$, normalized by the total speed $v=\sqrt{v_x^2+v_y^2+v_z^2}$. 
We use a mass resolution of $600 ~ \textrm{M}_\odot$ and $10^5$ particles for each UDG.

%
%

 %
 
 We model the Coma cluster in which the UDGs evolve through an analytic density profile representing the dynamical mass of the cluster in MOND, stemming from hydrostatic equilibrium of the X-ray emitting gas. 
 MOND has long been known to underpredict the deviation from GR needed to explain observations on galaxy cluster scales \citep[e.g.,][]{Sanders_1999, Sanders_2003, Angus_2008, Bilek2019a}, possibly implying a residual missing mass in clusters. We stress that the current cluster model includes both the baryonic component and this residual missing mass.
%
%
The density, $\rho_{\rm ana}$, is assumed to be spherically symmetric and  computed following \citet{Reiprich_2001} and \citet{Sanders_2003}, as explicated in \citet[][Section 4.1]{Freundlich_2022}. To highlight the importance of this analytic profile, the Coma cluster was also modelled using $10^6$ static particles distributed in spherical symmetry, which caused spurious dissolution of the UDGs upon close encounter with the cluster particles. It is to avoid this effect that a new patch, implementing an analytic density profile of the Coma cluster, was developed within the \textsc{por} context\footnote{This patch implements an analytic density profile of the Coma cluster in both MOND and Newtonian framework, and is available here:
\url{github.com/SrikanthNagesh/Coma_analytic_density_PoR}\label{github_patch}}. Similarly, \citet{Candlish_2018} have implemented analytic density profiles for the Coma and Virgo clusters in the \textsc{raymond} context.
In the MOND framework, the gravitational field within the UDG is a combination of the external field from the galaxy cluster and the self-gravity of the UDG. So, for a UDG at equilibrium within the cluster, before being potentially heated up by tides, one should take into account the external field from the cluster when generating the initial conditions for the UDG. Several analytic approximations exist for this \citep{Famaey_McGaugh_2012,Famaey_2018, Muller_2019, Haghi_2019}, and we choose here the one proposed by \citet[][Sect. 4.2, Eq. 25, cf. also \citealt{Oria_2021}]{Freundlich_2022}.

We run two sets of simulations: 
\begin{enumerate}
    \item To test whether cluster tides can heat up UDGs in the MOND context, we first run 36 simulations with UDGs placed at distances $R_i = 1, 1.2, 1.4, 1.6, 1.8, \rm{and} ~2.0$ Mpc from the cluster centre, respectively, and at each $R_i$, the UDGs are launched on orbits with different eccentricities $e = 0, 0.2, 0.4, 0.6, 0.8, 0.99$. Their velocity components are set to be $v_x = v_c e$, and $v_y = v_c \sqrt{1-e^2}$, where $v_c$ is the MOND circular velocity of the galaxy cluster at a given $R_i$; and
    they are not launched at apocenter, but rather at fixed distances $R_i$ in the direction of the cluster center. For this set of simulations, a box-size of 8 Mpc, $level\_min = 8 $, and a $level\_max = 15$ is used. The $level\_min$ sets the size of coarse grid cell, and $level\_max$ sets the size of the maximum resolved grid cell, given as $box\_length/2^{lmax}$, which in our case is 244 pc. The simulated UDGs are advanced for 5 Gyr with 100 Myr time-intervals. To check for the adequacy of the chosen resolution, we also ran a few comparison simulations by doubling the spatial resolutions (most resolved grid cell of 122~pc) and increasing the mass resolution (and number of particles) by a factor 10. The results were the same within 1\%, justifying our resolution choice.\\ 
    \item  Then, to test whether UDGs on a first infall could keep the memory of their velocity dispersion in isolation, an additional set of simulations is then run with $R_i$ varying from $10~\rm{Mpc}$ to  $14~\rm{Mpc}$ and $e= 0.99$.
    In this second set of simulations, a box-size of 22 Mpc, and $level\_max = $ 16 is used.
    These simulations are run for 7 Gyr with outputs at 100 Myr interval. 
    The positions, velocities, and mass of the UDG particles are extracted from the output of \textsc{por} using the \textsc{extract\_por} software \citep{Nagesh_2021}. 
\end{enumerate}

\section{Results}
\label{section:Results}

\subsection{Heating by tides?}
\label{subsection:tides}

\begin{figure}
    \includegraphics[width = 0.49\textwidth]{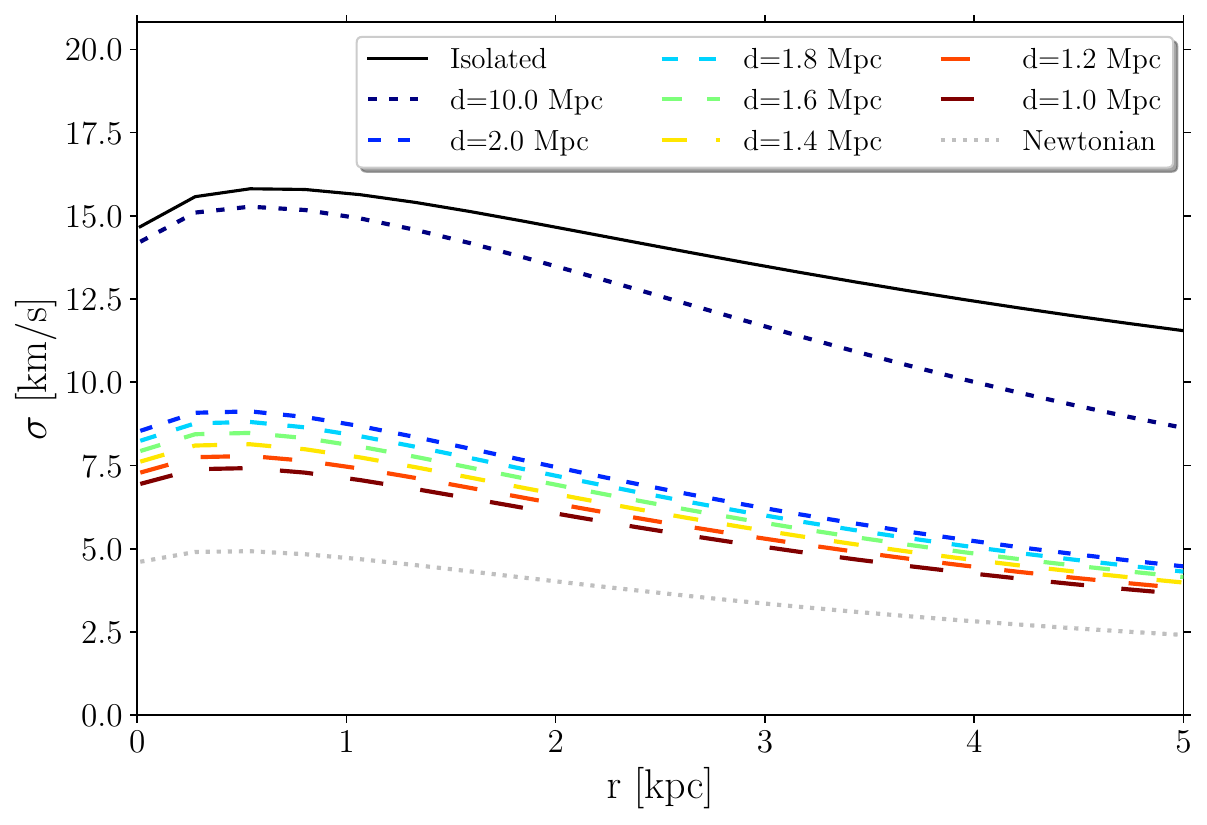}
    \caption{Initial equilibrium velocity dispersion profile of the simulated UDG, computed by solving Jeans equation and taking into account the external field at the launch radius. 
    For comparison, the solid black line corresponds to MOND in isolation while the dotted light gray line shows the Newtonian prediction. The EFE decreases the velocity dispersion from the isolated MOND prediction, making the profile closer to the Newtonian prediction. 
    }
    \label{fig:UDG_sigma_los_theoretical}
\end{figure}

\begin{figure}
    \includegraphics[width = 0.49\textwidth]{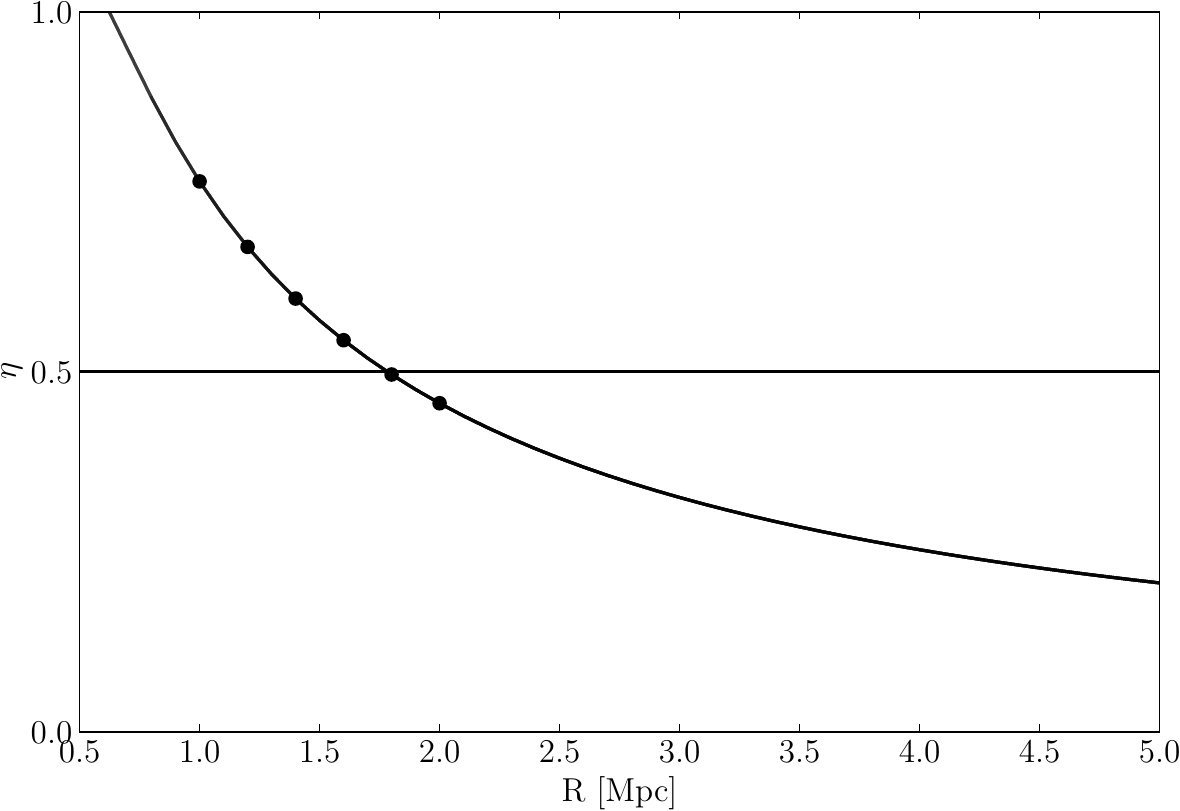}
    \caption{Tidal susceptibility $\eta$ derived from Eq.~(\ref{eq:tidal_susc}) as a function of distance from the cluster centre. The solid black points mark the tidal susceptibility at the distances where the UDGs were launched. Points above the horizontal $\eta=0.5$ line are expected to be at least partially affected by tides.}
    \label{fig:tidal_suscp}
\end{figure}

\begin{figure*}
    \includegraphics[width = 1.0\textwidth,clip]{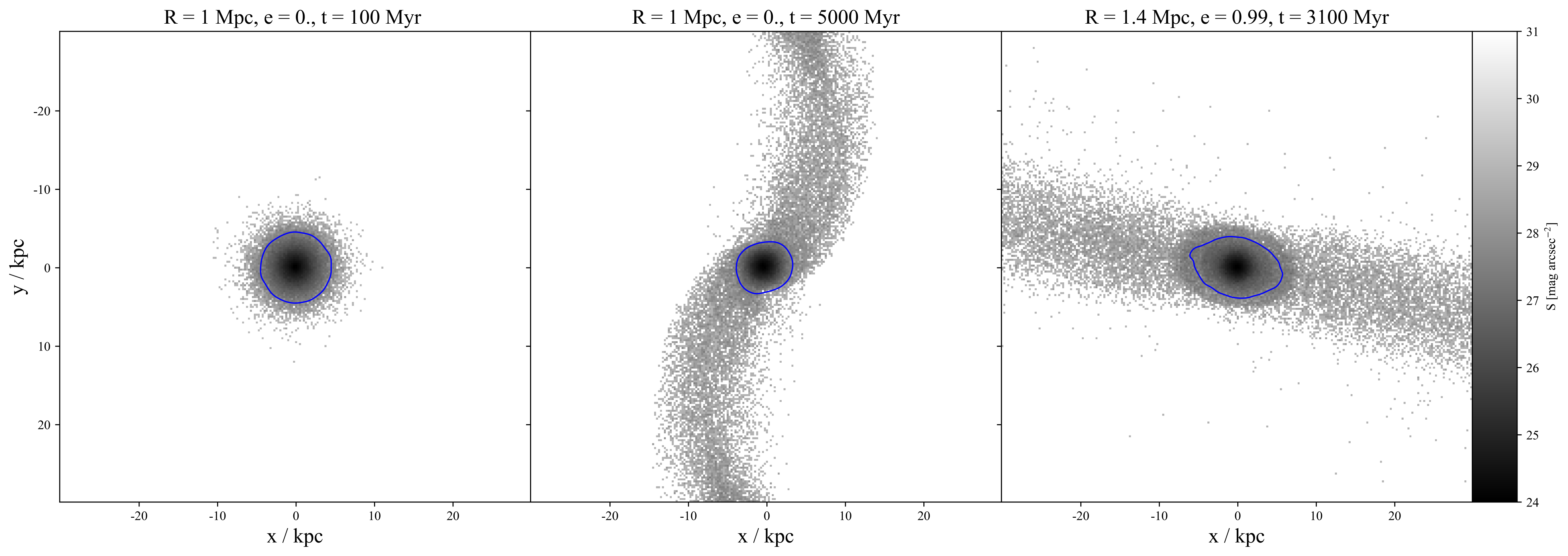}
    \caption{Projected surface density maps of simulated UDGs. \textit{Left:} UDG launched from $R = 1 ~\rm Mpc$ with an eccentricity $e =  0 $  after 0.1 Gyr. \textit{Middle:} Same UDG after 5 Gyr, with tidal tails. \textit{Right:} UDG launched from $R = 1.4 ~\rm Mpc$ with $e =  0.99$ after 3.1 Gyr. In all panels, the blue contour corresponds to a surface brightness threshold of $29.5~\rm  mag ~arcsec^{-2}$. 
    }
    \label{fig:shells}
\end{figure*}

\begin{figure*}
    \centering
    \includegraphics[width=0.49\textwidth]{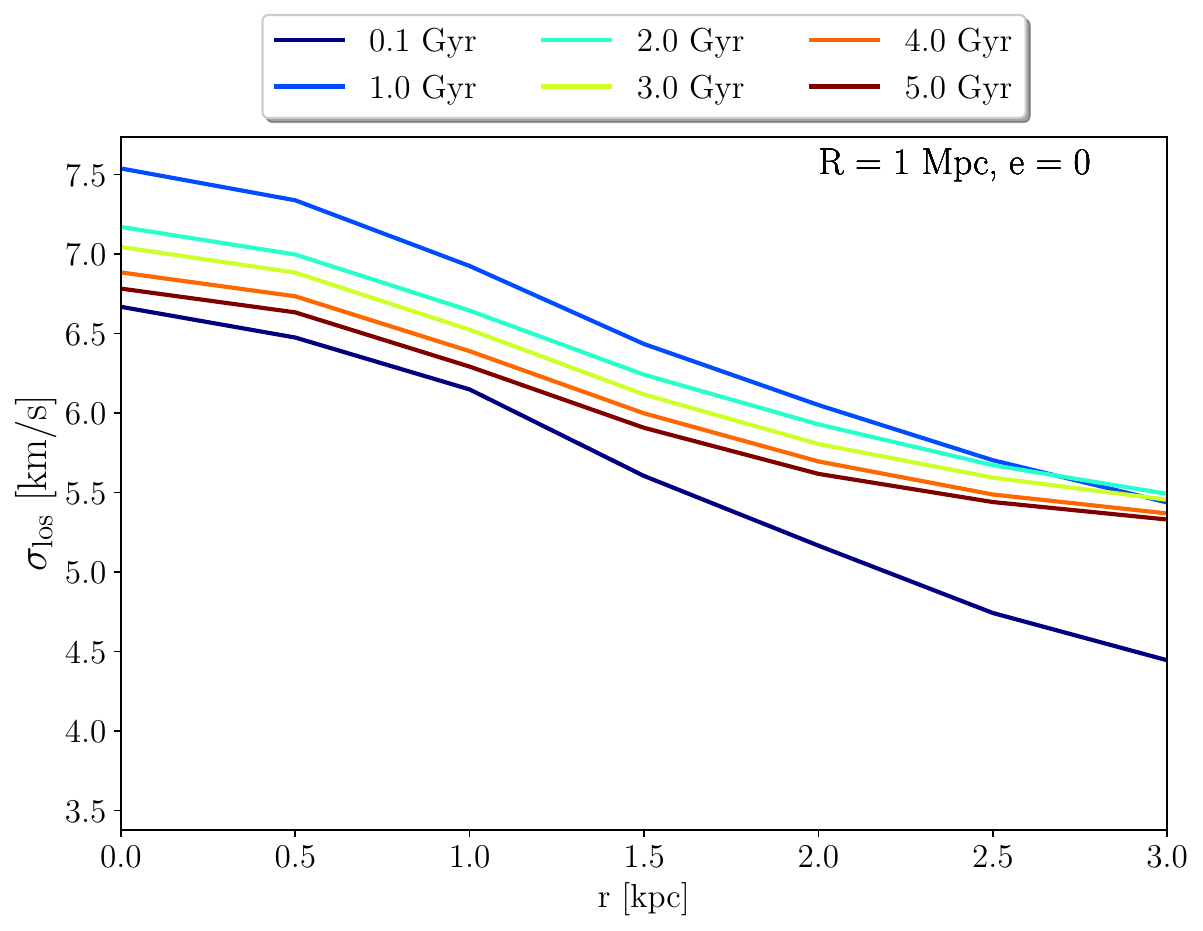}
    \hfill
    \includegraphics[width=0.49\textwidth]{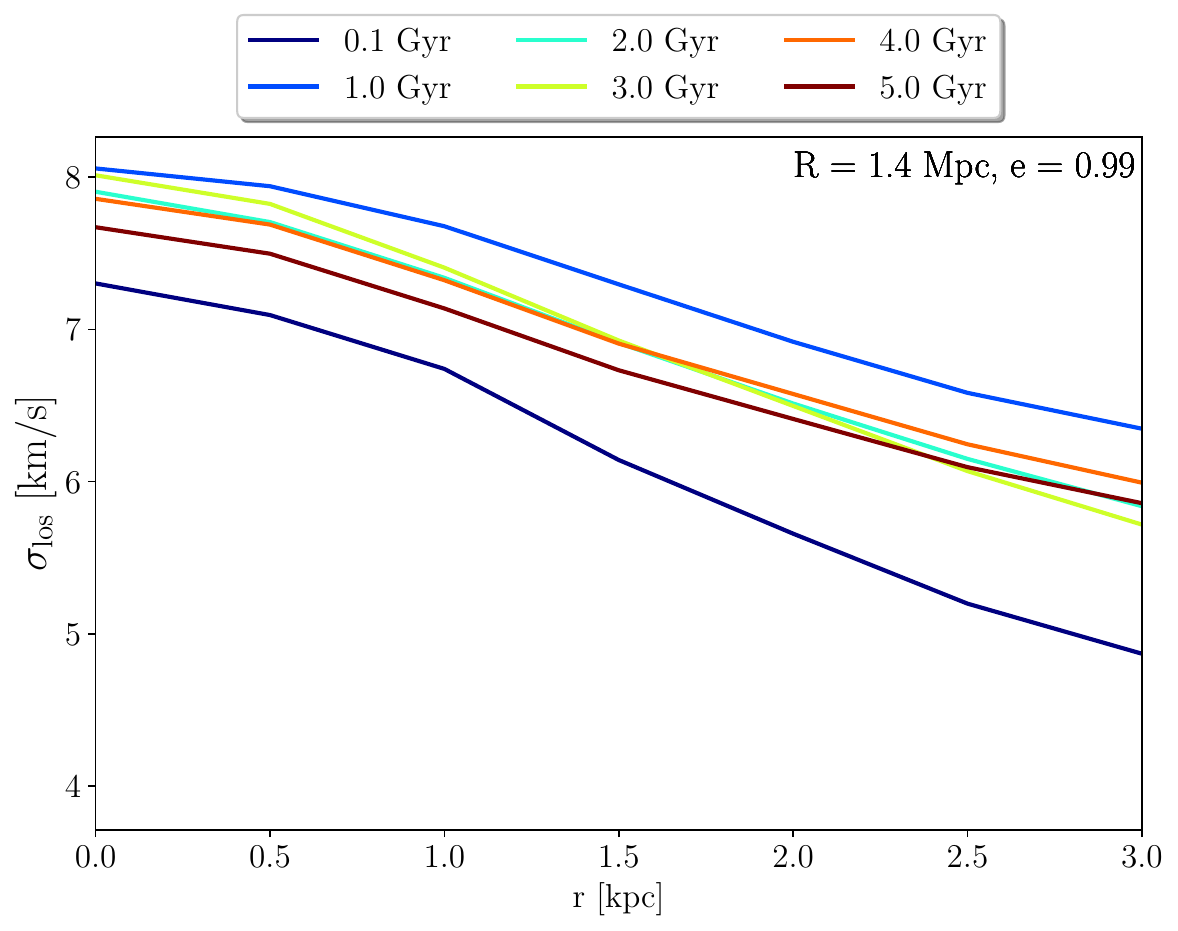}

   \caption{Evolution of the los velocity dispersion ($\sigma_{\rm los}$) of the two simulated UDGs shown in Fig.~\ref{fig:shells}. {\it Left:} UDG launched from $R = 1 ~\rm Mpc$ on a circular orbit with an eccentricity $e =  0 $. \textit{Right:} UDG launched from $R = 1.4 ~\rm Mpc$ on a radial orbit with $e =  0.99$.
   }
   %
   \label{fig:sigma_los_fig3}
\end{figure*}

\begin{figure*}
    \includegraphics[width = 1.0\textwidth]{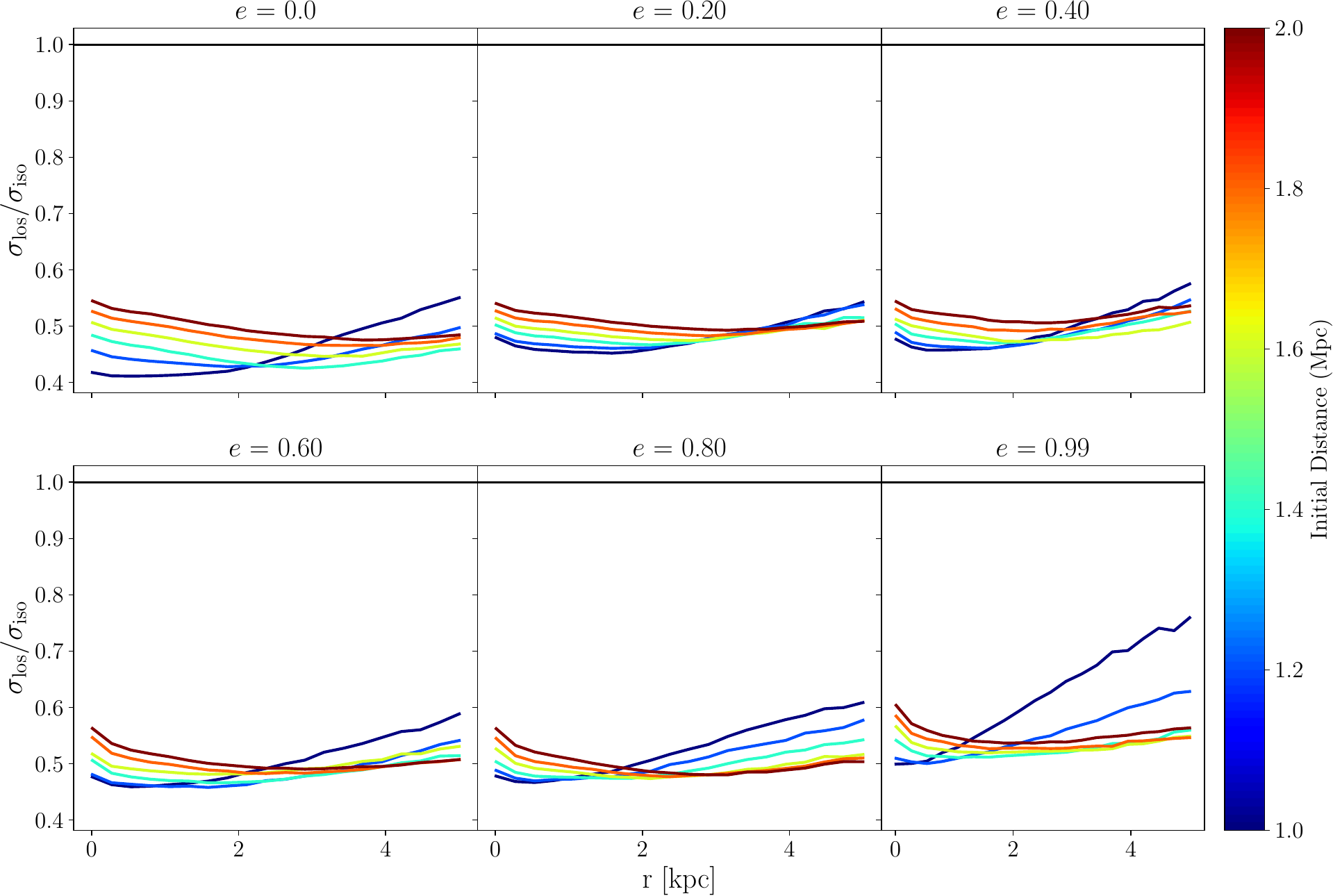}
    \caption{Ratio between the line-of-sight velocity dispersion ($\sigma_{\rm los}$) of the simulated UDGs at apocenter and the corresponding expected isolated MOND prediction ($\sigma_{\rm iso}$) as a function of radius. The six panels are arranged in the order of increasing eccentricity of the orbits of the launched UDGs, and the six different curves are colored based on the initial launch distance of the UDGs. 
    The horizontal black line corresponds to $\sigma_{\rm los} = \sigma_{\rm iso}$. The stellar velocity dispersion of observed Coma cluster UDGs are generally within $30\%$ of $\sigma_{\rm los}$ given their uncertainties (cf. Table~1 and Fig.~4 of \citealp{Freundlich_2022}), a regime that no simulated UDG reaches.
    %
    } 
    \label{fig:vel_disp_tile}
\end{figure*}

\begin{figure*}
    \includegraphics[width = 1.0\textwidth]{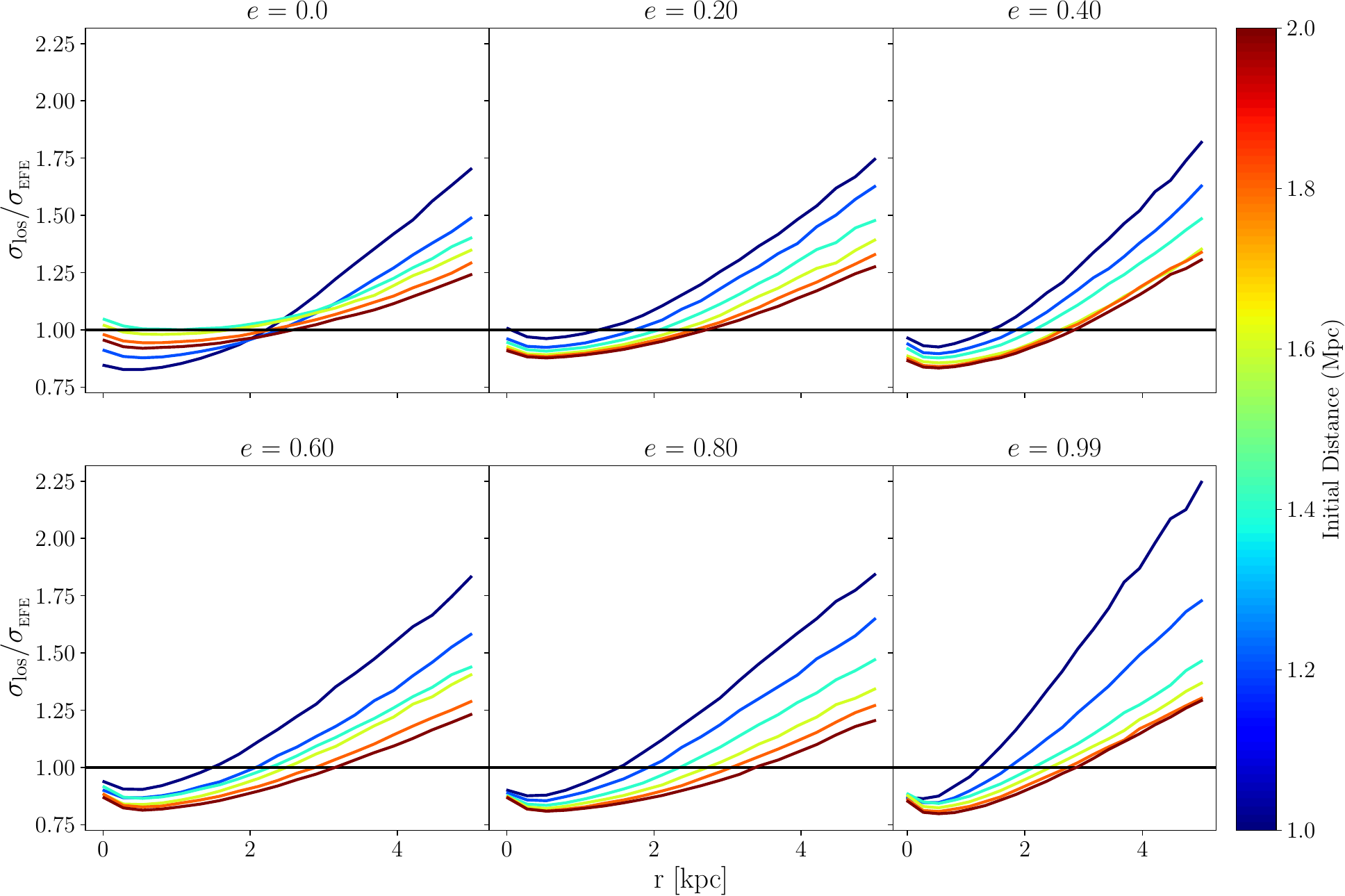}
    \caption{Same as Fig.~\ref{fig:vel_disp_tile} but for the ratio between $\sigma_{\rm los}$ and $\sigma_{\rm EFE}$ the Jeans equilibrium MOND prediction with EFE. The outskirts of the simulated UDGs are affected by tides but not enough to reach $\sigma_{\rm iso}$.}
    \label{fig:vel_disp_EFE_tile}
\end{figure*}

For our first set of simulations with different initial radii and eccentricities, Fig.~\ref{fig:UDG_sigma_los_theoretical} shows the equilibrium velocity dispersion profiles at launch, taking into account the EFE at the initial radius with Eq. 25 of \citet[][]{Freundlich_2022}. It can be seen clearly that the MOND EFE due to the cluster potential lowers the self-gravity of UDGs which renders them more susceptible to tidal forces \citep{Brada_2000, Asencio_2022} that could play an important role in enhancing the velocity dispersion of these systems \footnote{For a UDG on a circular orbit at 10~Mpc from the cluster center, where the tidal heating is negligible, the velocity dispersions remain very close to the initial setup after launching the simulation, thereby validating the adopted EFE formula at equilibrium}. Signatures of tidal interactions like tidal streams \citep{Mihos2015, Wittmann2017, Bennet2018}, elongation \citep{Koch_2012, Merritt2016, Toloba2016, Venhola2017, Lim2020}, and gas kinematics \citep{Scott2021}, have indeed been observed in UDGs of the Coma cluster.

In order to understand the effect of tides, we first calculate the tidal susceptibility 
\begin{equation}
    \eta = \dfrac{r_{_{1/2}}}{r_{2}}, 
    \label{eq:tidal_susc}
\end{equation}
where ${r_{_{1/2}}} \approx (4/3) R_e$ is the de-projected half mass radius and $r_{2}$ is the Roche lobe radius  perpendicularly to the axis linking the centre of the cluster and the UDG. The latter radius $r_2$ is calculated using the inner Lagrange point $r_1$, itself obtained by numerically solving the equation equating the UDG internal gravity with the tidal force (see Appendix B of \citealt{Freundlich_2022} for a complete derivation). 
For simulated UDGs as a function of distance from the centre of the cluster, Fig.~\ref{fig:tidal_suscp} shows $\eta$, which is significant at small cluster-centric radii (< 2Mpc).

In \citet{Freundlich_2022}, the measured line-of-sight (los) velocity dispersions of Coma cluster UDGs were found to be in relatively good agreement with the MOND prediction in isolation. Here, we therefore compare the los (along the $z$-axis of the cluster) velocity dispersions $\sigma_{\rm los}$ of the simulated UDGs with the prediction of MOND in isolation.
After extracting all the output particle data using \textsc{extract\_por}, we subtract the barycentre in position and velocities of all the particles at each snapshot. This subtraction allows one to calculate quantities in the rest frame of the UDGs, and reduces the effect of numerical drift. 
At each snapshot, we construct annuli of 0.25 kpc up to a projected radius of 5~kpc, which corresponds approximately to the distance of the furthest velocity dispersion measurement in DF44 \citep{vanDokkum2019b}. We however note that the typical radius in the observed sample within which we have data for most UDGs of the Coma cluster is of the order of 2~kpc or less. The los velocity dispersion is calculated in each bin with the unbiased estimator of the standard deviation

\begin{equation}
    \sigma_{\rm los}=\sqrt{\frac{   \sum\limits_{i=1}^{N}  v_{los, i}^2 - \dfrac{1}{N} \left(  \sum\limits_{i=1}^N  v_{los, i} \right)^2}{N-1}}
    \label{eq:velocity-dispersion}
\end{equation}
%
where $N$ is the total number of particles in a given bin.
For the UDGs to completely experience the effect of tides, we let them make at least one pericentric passage on their orbit, and consider the UDGs at apocentres since they spend a longer time close to apocentre than to pericentre, except for the case of circular orbits where we analyse the last snapshot of the simulation. 

The theoretical los velocity dispersion in the MOND context is derived from S\'ersic fits of the surface density maps in the ($x,y$) plane using \textsc{galfit} \citep{Galfit_software}, in order to emulate observational studies such as \citet{Freundlich_2022}. To compute mass throughout our simulations, we generate surface density maps along the (los) $z$-axis of the cluster at each snapshot, we compute isodensity contours corresponding to a g-band surface brightness of 29.5 mag arcsec$^{-2}$, 
in surface brightness using $M = M_{g}$ + 21.572 - 2.5 log$_{10} (L_\odot/$pc$^2)$, where $M_g$, and $L_\odot$ are the absolute g-band magnitude and luminosity of the Sun, 
and we adopt a mass-to-light ratio of 1 to convert luminosity into mass.
The 29.5 mag arcsec$^{-2}$ is motivated by possible upcoming surveys with the Euclid Visual instrument \citep{Euclid_LSB}.
It is important to note that the mass of the UDG varies along the simulation, since we only consider the mass enclosed within the 29.5 surface brightness contour. Figs.~\ref{fig:shells} and \ref{fig:sigma_los_fig3} illustrate the typical evolution of UDGs along the simulation in terms of projected surface density maps and los velocity dispersion.
%
%
We note that in all surface density maps, the observable parts of the galaxies within the  $29.5~\rm  mag ~arcsec^{-2}$ threshold look relatively relaxed, implying that that our current observational view of such galaxies may be limited due to sensitivity but that future deeper images may reveal that the outer parts of UDGs are severely tidally distorted.
We deproject the two-dimensional best-fit S\'ersic profile as indicated in Section~\ref{sec:methods}, using the semi-analytical approximation by \citet{Lima_Neto1999} and the update by \citet{Marquez2000}, compute the radial velocity dispersion from the Jeans equation assuming isotropy, and convert it into the expected los velocity dispersion by projecting the velocity ellipsoid along the los (cf. \citealp{Freundlich_2022}, Section 3.2.1, as well as \citealp{Binney1982} and \citealp{Mamon2005}).

We derive the theoretical los velocity dispersion both in isolation, hereafter $\sigma_{\rm iso}$, and in the presence of an external field, $\sigma_{\rm EFE}$
. Fig. \ref{fig:vel_disp_tile} shows the ratio of the simulated $\sigma_{\rm los}$, measured with Eq.~(\ref{eq:velocity-dispersion}), 
to the isolated MOND theoretical prediction $\sigma_{\rm iso}$. 
The main result here is that the $\sigma_{\rm los}$ profiles of UDGs on {\it all} orbits remain significantly below the isolated MOND prediction indicating that tides do not increase $\sigma_{\rm los}$ from their initial equilibrium values (within the external field from the cluster) up to the isolated MOND prediction. Nevertheless, UDGs launched from  $1$ and $1.2$ Mpc have rising $\sigma_{\rm los}$ profiles that become
steeper as a function of eccentricity, indicating that the outskirts are significantly affected by tides, but not enough to reach the isolated MOND prediction. 
To check how much heating is produced with respect to the equilibrium prediction within the external field of the cluster, $\sigma_{_{\rm EFE}}$, the ratio $\sigma_{\rm los}/\sigma_{_{\rm EFE}}$ is plotted in Fig.~\ref{fig:vel_disp_EFE_tile}. In all cases, the latter ratio in the the inner parts of the UDG remains close to 1, especially for low-eccentricity orbits (this also justifies {\it a posteriori} our chosen analytical prescription for the internal gravitational field in the presence of an EFE), confirming that the inner part of the UDG is in equilibrium within the EFE, while the outer parts are not, due to tides, although not enough compared to observations.

\subsection{First infall onto the cluster?}
\label{subsection:recent_infall}

\begin{figure*}
    \includegraphics[width = 1.0\textwidth,clip]{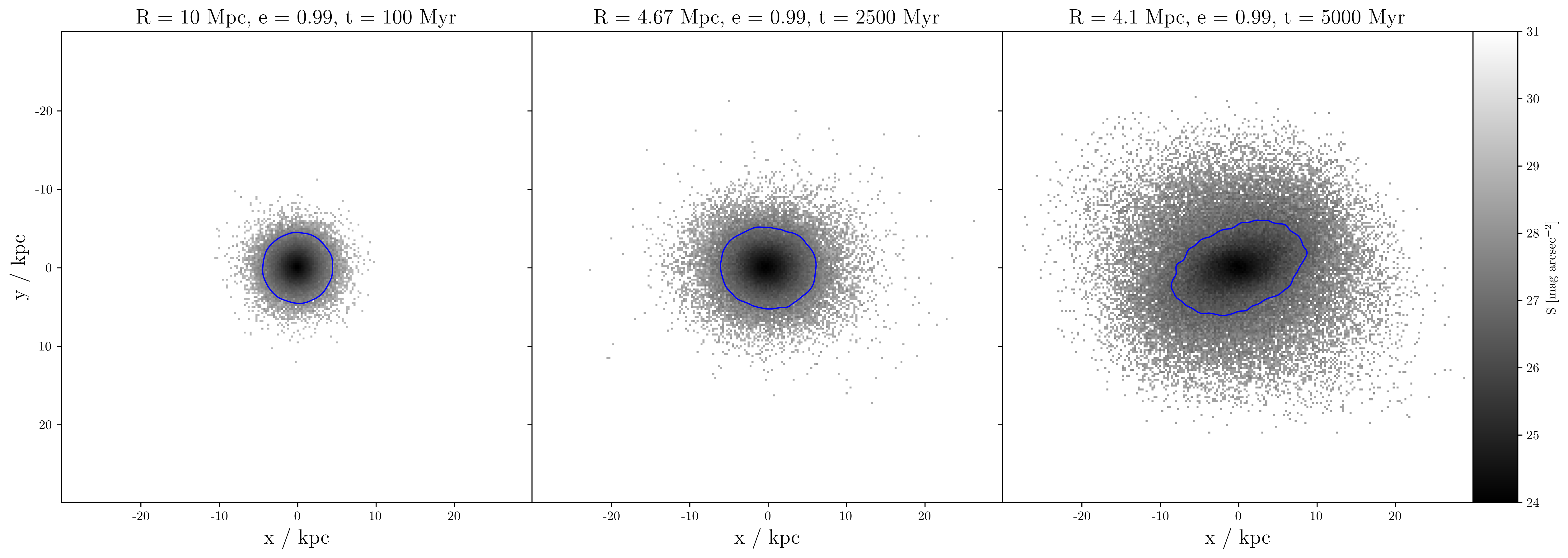}
    \caption{Projected surface density maps of a UDG launched from $R = 10 ~\rm Mpc$ on a radial orbit with an eccentricity $e =  0.99$ at different times. As in Fig.~\ref{fig:shells}, the blue contours correspond to a surface brightness threshold of $29.5~\rm  mag ~arcsec^{-2}$. 
    %
    }
    \label{fig:10Mpc}
\end{figure*}

\begin{figure*}
    \centering
    \includegraphics[width=0.49\textwidth]{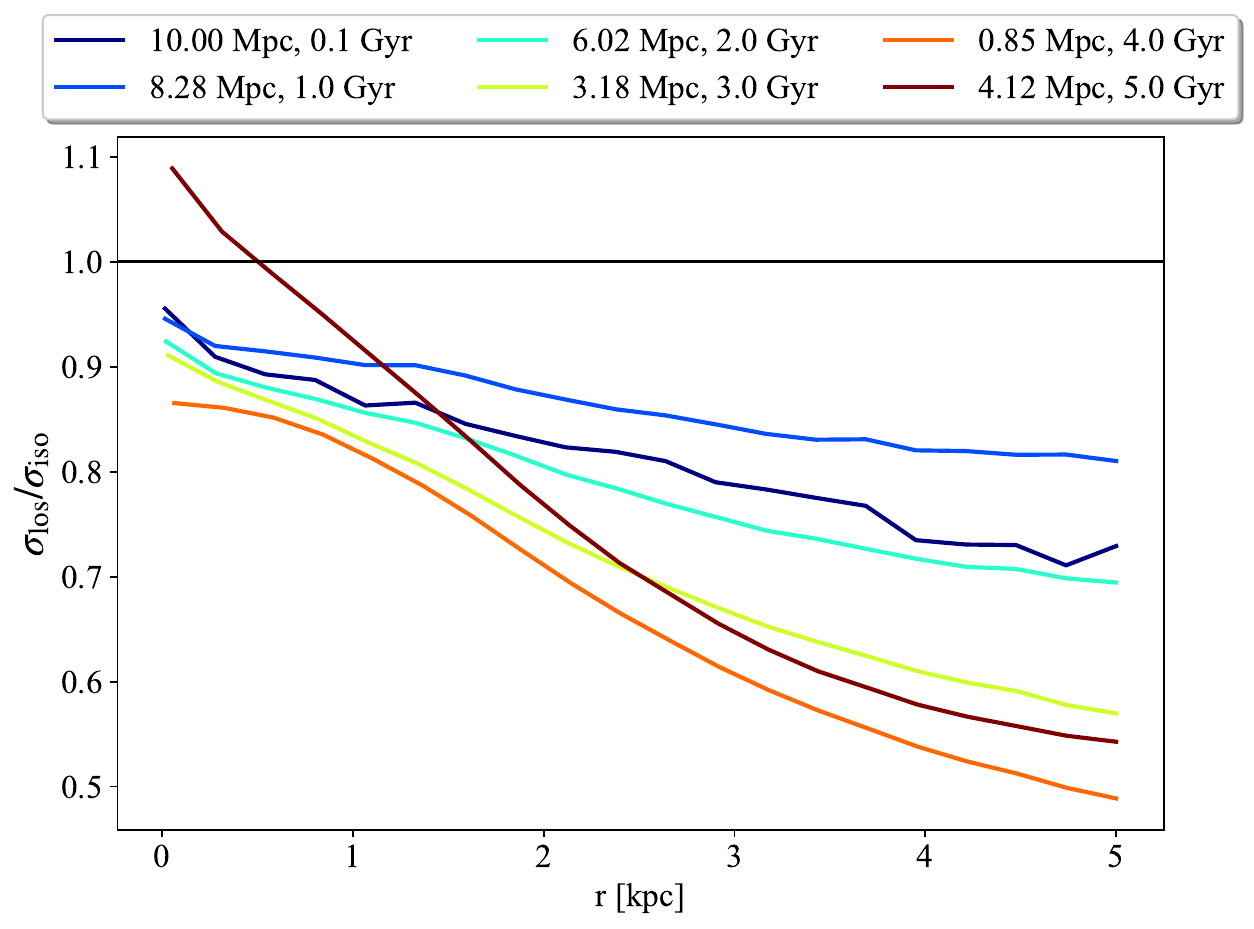}
    \hfill
    \includegraphics[width=0.49\textwidth]{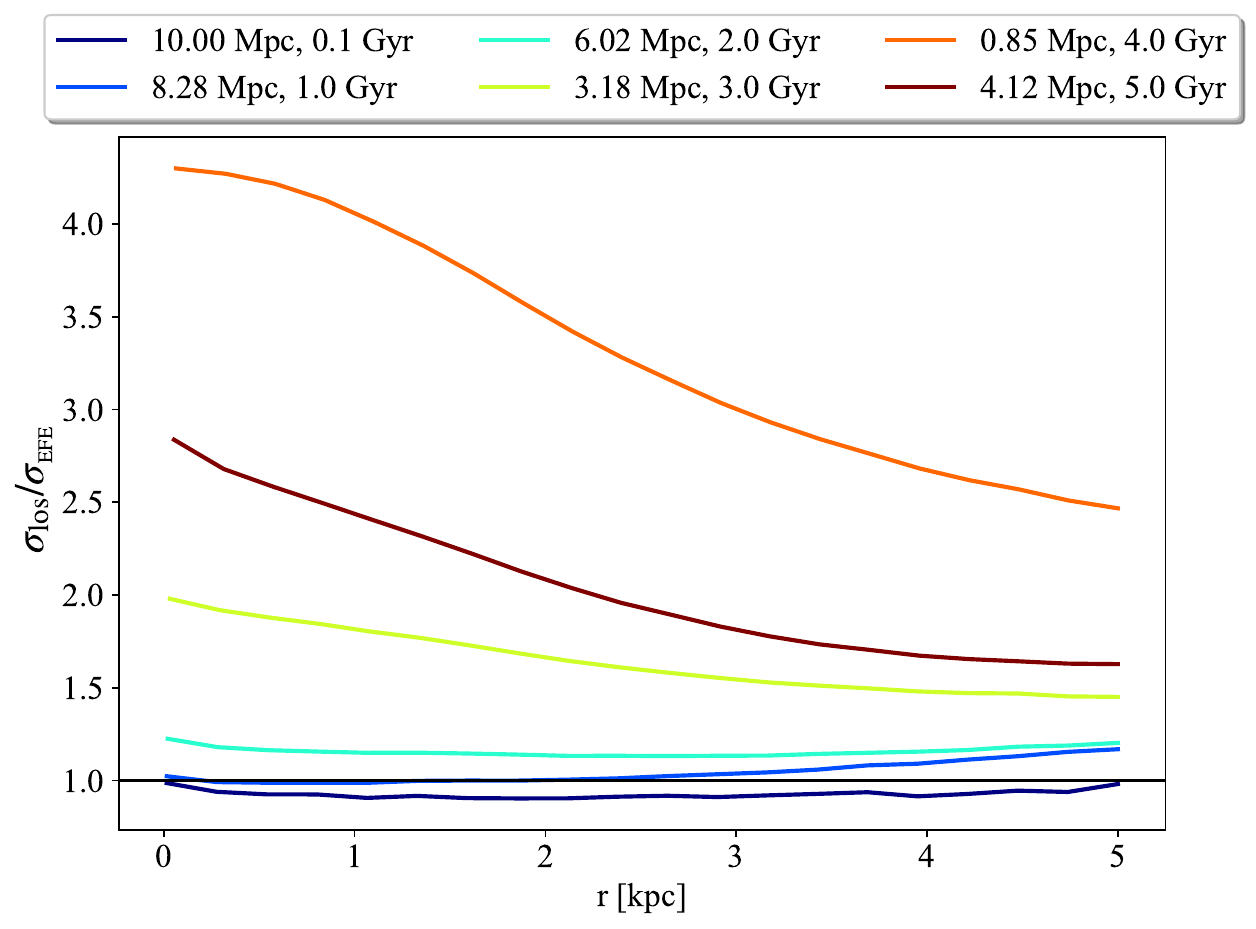}

   \caption{Evolution of the ratios between the los velocity dispersion ($\sigma_{\rm los}$) and that predicted by MOND in isolation ($\sigma_{\rm iso}$, {\it left panel}) and including the EFE ($\sigma_{_{\rm EFE}}$, {\it right panel}) for a simulated UDG launched on a radial orbit ($e=0.99$) from 10 Mpc from the cluster center. The UDG reaches pericenter around 4 Gyr. Its velocity dispersion decreases until pericenter passage (cf. $\sigma_{\rm los}/\sigma_{\rm iso}$), however without equilibrating with the EFE (cf. $\sigma_{\rm los}/\sigma_{_{\rm EFE}}$), and it undergoes tidal heading after pericenter passage. Numbers in the legend indicate the simulation time and the distance from the cluster center.
   }
   \label{fig:sig_iso_10Mpc}
\end{figure*}



Our second set of simulations of UDGs launched from 10 to 14 Mpc on radial orbits is meant to test whether UDGs on their first infall may not have time to equilibrate themselves with the EFE and could thus retain the velocity dispersion they had in isolation. As can be seen in Fig.~\ref{fig:UDG_sigma_los_theoretical}, the velocity dispersion in equilibrium at 10 Mpc is indeed close to that expected in isolation, reaching 90\% of the isolated MOND velocity dispersion in the central parts. 
Since it takes $\sim 6$ Gyr for a UDG to fall towards the central 3 Mpc, we estimate that $\sim 166$ UDGs have to be accreted per Gyr to reach the number of observed UDG candidates in the Coma cluster \citep[$\sim 10^3$][]{Bautista_2023, Zaritsky2019,Yagi_2016, Koda_2015}.
%
 Fig.~\ref{fig:10Mpc} shows the evolution of one such UDG, while Fig.~\ref{fig:sig_iso_10Mpc} shows the ratio of $\sigma_{\rm los}/\sigma_{\rm iso}$ and $\sigma_{\rm los}/\sigma_{_{\rm EFE}}$ at different times for a UDG launched from 10 Mpc on a radial orbit with an eccentricity $e=0.99$. 
It shows that from launch to pericenter the velocity dispersion decreases, especially towards the outskirts, but not sufficiently to equilibrate with the EFE : close to pericenter, the velocity dispersion reaches more than 4 times its equilibrium value under the EFE. 
After pericenter the UDG undergoes tidal heating with an increase of the velocity dispersion, especially at its center, and it starts to equilibrate with the EFE in the outskirts. We note that this central increase of the velocity dispersion may be precisely in line with some of the velocity dispersion profiles of Coma cluster UDGs reported by \citet[][cf. also \citealp{Freundlich_2022}, Fig. 4]{Chilingarian2019}. 
Consequently, UDGs on their first infall could retain the relatively high velocity dispersion they had in isolation, but they equilibrate with the EFE after pericenter passage. 

\begin{figure}
    \includegraphics[width = 0.49\textwidth]{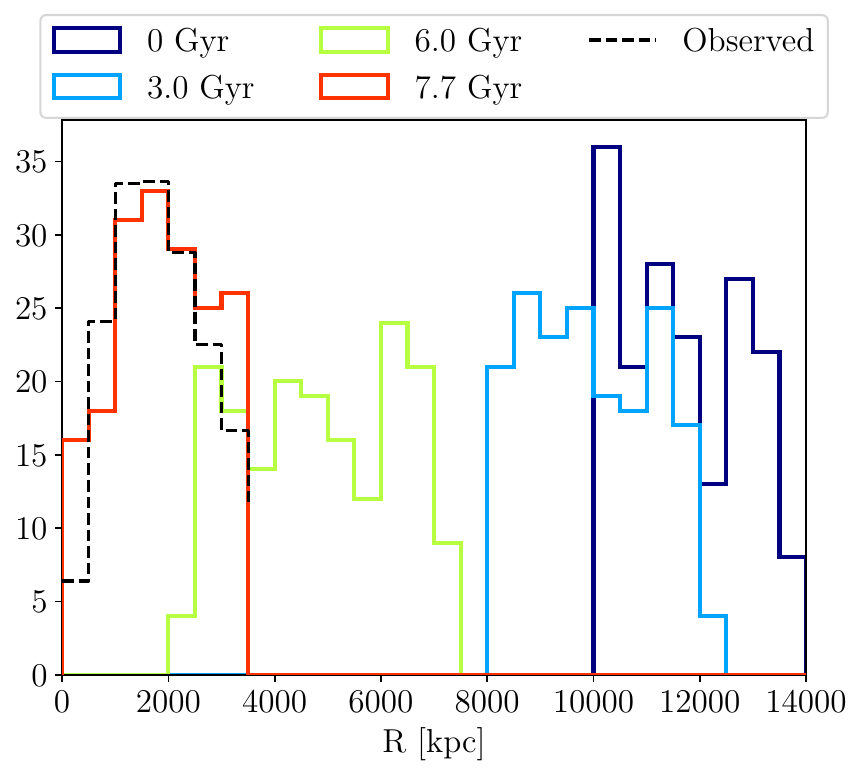}
    \caption{
    Evolution of a UDG population launched between 10 and 14 Mpc with an inward velocity of 100 km/s which recovers approximately the observed UDG distribution after 7.7 Gyr. 
    }
    \label{fig:initial_distribution}
\end{figure}

If cluster UDGs are on their first infall, their observed distribution places relatively strong constraints on their assembly history. 
%
A possibility would be that they fell together onto the cluster along cosmic filaments, as suggested by several observations \citep{Van_Dokkum_2019b, Zaritsky2019}.
%
To illustrate this scenario, we consider a population of UDGs falling onto the cluster from 10-14 Mpc with an initial inward radial velocity of 100 km/s, which corresponds to the lower bound of average kinetic bulk flow velocities in cosmic filaments \citep[e.g.][]{Kraljic_2019}, and follow this population as it falls towards the cluster center. Fig.~\ref{fig:initial_distribution} displays the evolution of this UDG population, whose initial distribution was specifically chosen such that its final distribution would be comparable to the observed one. 
It shows that, with an initial cylindrical density distribution within filaments that would be almost flat, with a slight increase towards the cluster center, such an accretion event almost 8~Gyr ago would allow to recover a distribution very similar to that of the observed Coma cluster UDGs, notably if they came together from a cosmic filament. However, the observed distribution of UDGs is isotropic and would require at least a few such filamentary accretions at a roughly similar time for this scenario to work. This radial infall scenario is therefore the only viable scenario for explaining the kinematics of UDGs in MOND, if they have baryonic masses as observationally estimated.

\section{Conclusion}
\label{sec:conclusion}

Ultra-diffuse galaxies (UDGs) in clusters provide a testing ground for modified Newtonian dynamics (MOND) and its external field effect (EFE) given their low internal gravitational acceleration and the presence of an external field. Previous work showed that the velocity dispersion of Coma cluster UDGs are in-line with the MOND prediction in isolation but in tension with the EFE \citep{Freundlich_2022}. This result may either contradict MOND or point towards a yet-to-be found theory underpinning the MOND phenomenology in which the EFE would be screened inside clusters. In the classical MOND context, the tension could however be alleviated if the Coma cluster UDGs had much higher baryonic mass than currently estimated, if these UDGs were heated by tides, or if they fell recently onto the cluster such that they retained part of the high velocity dispersion they had in isolation.

Here, we investigated the latter two possibilities by running $N$-body simulations of UDGs in a cluster potential using the phantom of ramses \citep[\textsc{por};][]{Lughausen_2015, Nagesh_2021} patch of the adaptive mesh refinement code \textsc{ramses} \citep{Teyssier_2002}. In order to eliminate spurious noise due to discrete cluster particles, we implemented an analytical external density within the \textsc{por} context for the first time, which is publicly available here$^{\ref{github_patch}}$. 
First, to test whether tides could heat up UDGs in the MOND context, we simulated UDGs on different orbits with initial radii between 1 and 2 Mpc from the cluster center. We show that if UDGs are initially at equilibrium within the cluster external field, tides are not sufficient to increase their velocity dispersions to values as high as those observed (Section~\ref{subsection:tides}).

Then, to test whether UDGs on first infall could retain their high velocity dispersion, we simulated UDGs falling on radial orbits towards the cluster center from distances of 10 to 14 Mpc. 
We show that such UDGs on their first radial infall onto the cluster may retain their high velocity dispersions without being destroyed until their first pericentric passage (Section~\ref{subsection:recent_infall}). 
Hence, without alterations such as a screening of the EFE in galaxy clusters or higher baryonic mass, UDGs must be out-of-equilibrium objects on their first infall onto the cluster in the MOND context. 

We stress that this work relies on different assumptions and simplifications.
In particular, we only considered tidal forces from a smooth cluster potential, while actual clusters host substructures and other galaxies that can also influence the dynamics of UDGs and contribute to increasing their velocity dispersion or to accelerate their disruption. 
We further assumed initial conditions where the galaxies were already ultra-diffuse, while UDGs can in principle form through tidal heating in clusters \citep[e.g.][]{Jiang2019} or ram-pressure striping of gas-rich dwarf \citep[e.g][]{Grishin_2021}. The expansion of the stellar distribution is however expected to be accompanied by a decrease of the velocity dispersion during the relaxation phase if energy is conserved. 
We relied on the QUMOND formalism, as implemented in \textsc{por}, and estimated the EFE with an analytical formula derived and tested in this context. 
Finally, we carried out pure $N$-body simulations without taking into account any possible gaseous component, including the effect of ram-pressure stripping in the radial infall scenario.
It is also important to note that our simulation setup does not consider Hubble expansion of the Universe, and mass growth of the Coma cluster with time, which can slow down the infall of the UDGs and moderate the EFE. Both effects however are not expected to affect the qualitative conclusions of the present work. Modelling this scenario would require to rely on a formalism such as the spherical top-hat collapse model of \citep{Malekjani_2009} or on cosmological simulations in the MOND context.
Finally, the framework developed in the current work could also be applied in the future to dwarf spheroidal satellites in the Local Group.

\begin{acknowledgements}
STN thanks Gary Mamon, Nicolas Martin, Raphael Errani, Francoise Combes, and Jin Koda for useful discussions. STN, JF, BF and RI acknowledge funding from the European Research Council (ERC) under the European Union's Horizon 2020 research and innovation program (grant agreement No.\ 834148). MB acknowledges partial support from UK Science and Technology Facilities Council grant ST/V000861/1 and hospitality from University of St Andrews during  the visit. OM is grateful to the
Swiss National Science Foundation for financial support under the grant number PZ00P2\_202104.
\end{acknowledgements}
  
%
%
%
%
%


\bibliographystyle{aa}
\bibliography{Freundlich2021_MOND,UDG_article}


\end{document}